\documentclass[aps,prd,preprint,groupedaddress,preprintnumbers,floatfix]{revtex4}
\usepackage{graphicx}
\usepackage{amsmath,amssymb}
\def\lsim{\raise0.3ex\hbox{$\;<$\kern-0.75em\raise-1.1ex\hbox{$\sim\;$}}}
\def\gsim{\raise0.3ex\hbox{$\;>$\kern-0.75em\raise-1.1ex\hbox{$\sim\;$}}}
\newcommand{\CL}   { C.L. }

\begin{document}
\preprint{IFIC/07-71}
\title{Constraining Nonstandard  Neutrino-Electron Interactions}

\author{J. Barranco} 
\email{jbarranc@fis.cinvestav.mx}
\altaffiliation{Present address: Instituto de F\'{\i}sica, Universidad
Nacional Aut{\'o}noma de M{\'e}xico, Apdo. Postal 20-364, 01000
M{\'e}xico D.F., \ Mexico}
\author{O. G. Miranda}
\email{Omar.Miranda@fis.cinvestav.mx}
\author{C. A. Moura}
\email{cadrega@fis.cinvestav.mx}
\altaffiliation[On leave from: ]
{Instituto de Fisica Gleb Wataghin - UNICAMP, Brazil}
\affiliation{Departamento de F\'{\i}sica, Centro de Investigaci{\'o}n y de
  Estudios Avanzados del IPN, Apdo. Postal 14-740 07000
  M\'exico, D.F., Mexico}
\author{J.W.F.Valle}
\email{valle@ific.uv.es}
\affiliation{  AHEP Group, Instituto de F\'{\i}sica Corpuscular --
  C.S.I.C./Universitat de Val{\`e}ncia \\
  Edificio Institutos de Paterna, Apt 22085, E--46071 Valencia, Spain}

\begin{abstract}
  We present a detailed analysis on non-standard neutrino interactions
  (NSI) with electrons including all muon and electron (anti)-neutrino
  data from existing accelerators and reactors, in conjunction with
  the ``neutrino counting'' data ($e^+ e^- \to \nu \bar\nu \gamma$)
  from the four LEP collaborations.
  First we perform a one-parameter-at-a-time analysis, showing how
  most constraints improve with respect to previous results reported
  in the
  literature. 
  We also present more robust results where the NSI parameters are
  allowed to vary freely in the analysis.
  We show the importance of combining LEP data with the other
  experiments in removing degeneracies in the global analysis
  constraining flavor-conserving NSI parameters which, at 90\% and
  95\%~C.L., must lie within unique allowed regions.
  Despite such improved constraints, there is still substantial room
  for improvement, posing a big challenge for upcoming experiments.
\end{abstract}
\maketitle

\section{Introduction}

The historic discovery of neutrino oscillations constitutes the first
evidence for physics beyond the Standard Model, and one would like to
know to which direction it is pointing.
Despite a pretty good knowledge of the neutrino oscillation mechanism
and the parameters involved \cite{Maltoni:2004ei}, the nature of the
mechanism generating masses and mixings remains as elusive as ever.
Neutrino mass models fall in various classes~\cite{Altarelli:2004za}
involving models where neutrinos get mass \textit{a la
seesaw}~\cite{Minkowski:1977sc,gell-mann:1980vs,yanagida:1979,mohapatra:1980ia,schechter:1980gr,schechter:1982cv,mohapatra:1981yp},
those where neutrinos acquire mass radiatively due to the presence of
extra Higgs
bosons~\cite{zee:1980ai,babu:1988ki,AristizabalSierra:2006gb}, and
hybrid models, like those based on low energy supersymmetry with
spontaneous (or bilinear) breaking of
R-parity~\cite{Diaz:1997xc,Hirsch:2000ef,abada:2001zh,Diaz:2003as,Hirsch:2004he}.
Starting with the seesaw~\cite{schechter:1980gr} all those mechanisms
carry with them modifications to the structure of the standard
electroweak neutral and charged currents.  The simplest of such
modifications in the low energy regime may be written in the usual
$V-A$ form, similar to the four-Fermi interaction, characterized by
some new coupling, which we generally call, in what follows,
Non-Standard Interactions (NSI).
Such interactions can arise in a variety of ways, e.~g., from the
exchange of Higgs and/or supersymmetric scalar bosons as well as a new
heavy gauge boson, such as a $Z'$. Non-standard interactions can
conserve flavor, their differences characterizing the violation of
weak universality, known as Non-Universal (NU) NSI. Alternatively,
they may violate flavor conservation, known as Flavor Changing (FC)
NSI.

On the other hand, current neutrino oscillation data, as inferred from
the solar~\cite{Miranda:2004nb,Guzzo:2004ue,Bergmann:1998rg} and
atmospheric~\cite{gonzalez-garcia:1998hj,Friedland:2004ah,Fornengo:2001pm}
neutrino experiments, leave significant room for the existence of
sub-leading effects induced by NSI.
In fact, NSI effects may be comparable to oscillation effects in solar
neutrino physics, where indeed a new degenerate solution is still
allowed~\cite{Miranda:2004nb}.
At this stage laboratory experiments both from accelerators and
reactors can play a crucial role, since the strongest sensitivity to
NSI comes from this kind of experiments~\cite{Davidson:2003ha}.
Precision measurements of oscillation parameters in long baseline
oscillation experiments such as neutrino factories will also benefit
from improved NSI studies. These could help resolving the confusion
between the two found in Refs.~\cite{huber:2001de,huber:2002bi} and
further discussed in Refs.~\cite{Kopp:2007rz,Ribeiro:2007ud}.

Here we address the current sensitivity on non-standard interactions
as inferred from a global analysis of processes involving
(anti)-neutrinos and electrons. 
Previous analyses have been performed in Refs.
\cite{Berezhiani:2001rs,Davidson:2003ha,Abdallah:2003np,Barranco:2005ps}. 
Our current analysis combines the relevant experimental ``neutrino
counting'' data from $e^+ + e^- \to \nu + \bar\nu + \gamma$ obtained
by the four LEP
collaborations~\cite{Barate:1997ue,Barate:1998ci,Heister:2002ut,Abreu:2000vk,Acciarri:1997dq,Acciarri:1998hb,Acciarri:1999kp,Ackerstaff:1997ze,Abbiendi:1998yu,Abbiendi:2000hh},
and summarized in Ref.~\cite{Hirsch:2002uv}, with all the 
$\nu_e + e \to \nu_e + e$ data obtained by
LSND~\cite{Auerbach:2001wg}, and the $\bar\nu_e + e \to \bar\nu_e + e$
interaction studied in reactor experiments, namely: Irvine
\cite{Reines:1976pv}, MUNU \cite{Daraktchieva:2003dr} and Rovno
\cite{Derbin:1993wy}. For the muon neutrino case the relevant
reactions are $\nu_\mu + e \to \nu_\mu + e$ and $\bar \nu_\mu + e \to
\bar \nu_\mu + e$, measured at CHARM II \cite{Vilain:1994qy}.
Our analysis is also novel in the sense that we adopt a model independent
approach, as general as possible, allowing simultaneous variations of
all NSI parameters.
In particular, we compare the restrictions obtained varying only one
parameter at a time, with those obtained in the case where all six
flavor-conserving parameters are left free.

The analysis sketched above is organized as follows: in
Sec. \ref{sec:nonst-inter-relev} the NSI are introduced and the
relevant cross sections are expressed as function of NSI parameters,
in Secs. \ref{sec:data} and \ref{sec:chi2-analysis} we briefly present
the data and the details of the $\chi^2$ analysis. The results are
presented in Sec. \ref{sec:results} and more discussion and outlook
are given in Sec. \ref{sec:discussion}.

\section{Non-standard interactions and relevant cross sections}
\label{sec:nonst-inter-relev}

Neutrino NSI constitute an unavoidable characteristic feature of gauge
models of neutrino mass, for example those where they arise from the
admixture of isodoublet and isosinglet neutral leptons, like models of
the generic seesaw type~\cite{schechter:1980gr}. Typically the masses
of the light neutrinos are obtained by diagonalizing the mass matrix
\begin{equation}
    \label{eq:SS}
    \begin{bmatrix}
        M_L & D \\
        D^T & M_R
    \end{bmatrix}
\end{equation}
in the basis $\nu,\nu^c$, where $D$ is the standard $SU(2) \otimes
U(1)$ breaking Dirac mass term, and $M_R = M_R^T$ is the large
isosinglet Majorana mass. In the absence of the isotriplet (type-II)
$M_L \nu\nu$ term~\cite{schechter:1980gr} the scheme is called type-I
seesaw~\cite{Minkowski:1977sc,gell-mann:1980vs,yanagida:1979,mohapatra:1980ia}.
In models with spontaneous breaking of lepton number symmetry one has $M_L
\propto 1/M_R$, a feature that comes from the study of the scalar
potential and holds both in the case of left-right models (gauged
lepton number)~\cite{mohapatra:1981yp} and the case of majoron models
(ungauged lepton number)~\cite{schechter:1982cv}.

The structure of the associated effective $SU(2) \otimes U(1)$ weak
$V-A$ currents is rather complex and deviates from
standard~\cite{schechter:1980gr}. The first point to notice is that
the heavy isosinglets will mix with the ordinary isodoublet neutrinos
in the charged current weak interaction. As a result, the mixing
matrix describing the charged leptonic weak interaction is a
rectangular matrix $K$~\cite{schechter:1980gr} which may be decomposed
as
\begin{equation}
    \label{eq:CC}
    K = (K_L, K_H)
\end{equation}
where $K_L$ and $K_H$ are $3 \times 3$ matrices. Note that the
  ``effective'' lepton mixing matrix $K_L$ relevant in oscillation
  studies is non-unitary~\cite{valle:1987gv}.  For papers addressing
  possible future tests of such non-unitary effects see for
  example~\cite{Goswami:2008mi} and references therein.
The corresponding neutral weak
  interactions are described by a non-trivial
  matrix~\cite{schechter:1980gr} $K^\dagger K$
\begin{equation}
    \label{eq:NC}
    \mathcal{L} = \frac{ig^\prime}{2 \sin\theta_W} Z_\mu \bar{\nu_L} 
      \gamma_\mu K^\dagger K \nu_L \, .
\end{equation}

Such structure of the charged and neutral weak currents provides an
origin for neutrino NSI. 
Note, however, that the smallness of neutrino mass, which follows due
to the seesaw mechanism $M_{\nu \: \mathrm{eff}} = M_L \, - \, D
M_R^{-1} D^T$ and the condition $ M_L \ll M_R \,,$ implies that,
barring fine-tuning, the magnitude of neutrino NSI and its effects are
expected to be negligible. 

However this need not be so in general. Since the number $m$ of $SU(2)
\otimes U(1)$ singlets is arbitrary, one may, for example, extend the
lepton sector of the $SU(2) \otimes U(1)$ theory by adding a set of
{\sl two} 2-component isosinglet neutral fermions, denoted ${\nu^c}_i$
and $S_i$, in each generation. In such $m = 6$ models one can consider
the $9 \times 9$ mass
matrix~\cite{mohapatra:1986bd,Gonzalez-Garcia:1989rw}
\begin{equation}
    \label{eq:MATmu}
    \begin{bmatrix}
        0 & D & 0 \\
        D^T & 0 & M \\
        0 & M^T & \mu
    \end{bmatrix}
\end{equation}
(in the basis $\nu, \nu^c, S$). The Majorana masses for the neutrinos are
determined from
\begin{equation}
    \label{eq:33}
    M_L = D M^{-1} \mu {M^T}^{-1} D^T \, .
\end{equation}

Since in the limit $\mu \to 0$ the exact lepton number symmetry is
recovered and neutrinos become massless~\cite{mohapatra:1986bd} this
scheme is sometimes called ``inverse seesaw''~\footnote{For
  supersymmetric version of the same model see
  Refs~\cite{Deppisch:2005zm,Deppisch:2004fa}. Attempts to embed
  extended seesaw schemes in SO(10) lead to even more exotic varieties
  of seesaw, like the linear seesaw described in
  Ref.~\cite{Malinsky:2005bi}.}.

This provides an elegant way to generate neutrino masses without a
super-heavy scale, the smallness of the neutrino mass indicated by the
oscillation interpretation of solar and atmospheric neutrino data is
ascribed to the smallness of $\mu$, which is natural in 't Hofft's
sense: the symmetry of the theory is enhanced in the limit of
vanishing $\mu$.
This automatically allows for a sizeable magnitude of neutrino NSI
strengths, unconstrained by the smallness of neutrino
masses~\footnote{It also provides an explicit example for flavour and
  CP violation completely dettached from the smallness of neutrino
  masses~\cite{bernabeu:1987gr,branco:1989bn,rius:1990gk}}.

The NSI which are engendered in this case will necessarily affect
neutrino propagation properties in matter, an effect that may be
resonant in certain
cases~\cite{valle:1987gv,nunokawa:1996tg,EstebanPretel:2007yu}.
They may also be large enough as to produce effects in the laboratory.

An alternative way to induce neutrino NSI is in the context of
low-energy supersymmetry without R-parity
conservation~\cite{Hall:1984id,Ross:1985yg,santamaria:1987uq} where
one may also have, in addition to
bilinear~\cite{Diaz:1997xc,Hirsch:2000ef,abada:2001zh,Diaz:2003as}
also trilinear $L$ violating couplings in the super-potential such as
\begin{gather}
    \label{eq:lq}
    \lambda_{ijk} L_i L_j E^c_k \, \\
    \lambda'_{ijk} L_i Q_j D^c_k
\end{gather}
where $L, Q, E^c$ and $D^c$ are (chiral) super-fields which contain
the usual lepton and quark $SU(2)$ doublets and singlets,
respectively, and $i,j,k$ are generation indices.
The couplings in Eq.~(\ref{eq:lq}) give rise at low energy to the
following four-fermion effective Lagrangian for neutrino interactions
with $d$-quark including
\begin{equation}
    \label{eq:effec}
   \mathcal{L}_\mathrm{eff}  =  - 2\sqrt{2} G_F \sum_{\alpha,\beta}
    \xi_{\alpha\beta} \: \bar{\nu}_{L\alpha} \gamma^{\mu} \nu_{L\beta} \:
    \bar{d}_{R}\gamma^{\mu}{d}_{R}\:\:\:\alpha,\beta = e,\mu, \tau \, ,
\end{equation}
where the parameters $\xi_{\alpha\beta}$ represent the strength of the
effective interactions normalized to the Fermi constant $G_F$.
One can identify explicitly, for example, the following {\sl non-standard}
flavor-conserving NSI couplings
\begin{align}
    \xi_{\mu\mu} &= \sum_j \frac{|\lambda'_{2j1}|^2}
    {4 \sqrt{2} G_F m^2_{\tilde{q}_{j L}}} \, , \\
    \xi_{\tau\tau}& = \sum_j \frac{|\lambda'_{3j1}|^2}
    {4 \sqrt{2} G_F m^2_{\tilde{q}_{j L}} }\, ,
\end{align}
and the FC coupling
\begin{equation}
    \xi_{\mu\tau} =  \sum_j \frac{ \lambda^\prime_{3j1} \lambda^\prime_{2j1} }
    {4\sqrt{2}G_F m^2_{\tilde{q}_{jL}} }
\end{equation}
where $m_{\tilde{q}_{j L}}$ are the masses of the exchanged squarks and $j =
1,2,3$ denotes $\tilde{d}_L, \tilde{s}_L, \tilde{b}_L$, respectively.
The existence of effective neutral current interactions contributing
to the neutrino scattering off $d$-quarks in matter, provides new
flavor-conserving as well as flavor-changing terms for the matter
potentials of neutrinos.  Such NSI are directly relevant for
solar~\cite{Miranda:2004nb,Guzzo:2004ue,Bergmann:1998rg} and
atmospheric neutrino
propagation~\cite{gonzalez-garcia:1998hj,Friedland:2004ah,Fornengo:2001pm}.

In what follows we consider a more general class of non-standard
interactions described via the effective four fermion Lagrangian,
\begin{equation}
-{\cal L}^{eff}_{\rm NSI} =
\varepsilon_{\alpha \beta}^{fP}{2\sqrt2 G_F} (\bar{\nu}_\alpha \gamma_\rho L \nu_\beta)
( \bar {f} \gamma^\rho P f ) \,,
\label{eq:efflag}
\end{equation}
where $G_F$ is the Fermi constant and $\varepsilon_{\alpha
  \beta}^{fP}$ parametrize the strength of the NSI.  For laboratory
experiments $f$ is a first generation SM fermion ($e, u$ or $d$). Here
we analyze only processes involving electrons, so that in what follows
we have only $f=e$.  The chiral projectors $P$ denote
$\{L,R=(1\pm\gamma^5)/2\}$, while $\alpha$ and $\beta$ denote the
three neutrino flavors: $e$, $\mu$ and $\tau$. 

In total there are 12 relevant parameters given by
$\varepsilon^{P}_{\alpha\beta}$. 
In order to constrain these we use experimental data reported by LEP
($e^+e^-\to\nu\bar\nu\gamma$), LSND ($\nu_ee\to\nu_ee$), CHARM II
($\nu_\mu e$ or $\bar\nu_\mu e$ scattering) and reactor experiments
($\bar\nu_ee\to\bar\nu_ee$). The cross sections for the interactions
of each experiment are given next.

\subsection{LEP cross section}
\label{sec:lep-cross-section}

The $e^+e^-\to\nu\bar\nu\gamma$ cross section can be calculated at
tree level using the `radiator' approximation to describe the photon
emission \cite{Nicrosini:1988hw} as
\begin{equation}
\sigma^{\rm theo}_{\rm LEP}(s) =  \int {\rm d} x  \int {\rm d} c_{\gamma}
~H(x,s_\gamma;s)~ \sigma_0^{\rm theo}(\hat{s}) \, ,
\label{eq:sm}
\end{equation}
where $s$ is the center of mass energy, $x=2E_\gamma/\sqrt2$,
$E_\gamma$ is the photon energy, $\sigma_0^{\rm theo}=\sigma_0^{\rm
  SM}+\sigma_0^{\rm NSI}$ is the cross section for the process
$e^+e^-\to\nu\bar\nu$ and $\hat s=(1-x)s$. Here the `radiator'
function $H$ is given by
\begin{equation}
H(x,s_\gamma;s)={2 \alpha \over \pi x s_\gamma }\left[
\left(1-{x \over 2}\right)^2+ {x^2 c_\gamma^2 \over 4} \right]\,,
\label{eq:rad}
\end{equation}
where $s_\gamma\equiv\sin\theta_\gamma$ and $c_\gamma^2\equiv1-s_\gamma^2$, with $\theta_\gamma$ being the photon emission angle.

Working in the limit of vanishing $W-\gamma$ interactions but
considering finite distance effects for the $W$ propagator, the
Standard Model $e^+e^-\to\nu\bar\nu$ cross section is given as
\begin{eqnarray}
\sigma_0^{\rm SM}(s) &= & \frac{N_\nu G_F^2}{6\pi} M^4_Z 
( g^2_{R} +g^2_{L}) \frac{s}{\left[ (s-M^2_Z)^2 +(M_Z \Gamma_Z)^2\right]}  \\
\nonumber
& + & \frac{G_F^2}{\pi} M^2_W\left\{ \frac{s + 2M^2_W}{2s} 
-\frac{M^2_W}{s} \left(\frac{s + M^2_W}{s}\right)
\log\left(\frac{s + M^2_W}{M^2_W}\right) \right . \\ \nonumber
& - & \left . g_{ L} \frac{M^2_Z (s-M^2_Z)}{(s-M^2_Z)^2 + (M_Z \Gamma_Z)^2}
\left[\frac{(s + M^2_W)^2}{s^2}
\log\left(\frac{s + M^2_W}{M^2_W}\right) - \frac{M^2_W}{s} -\frac{3}{2}\right]\right\},
\label{eq:smcross}
\end{eqnarray}
where $N_\nu$ is the number of neutrino families, $g_R$ and $g_L$ are
the SM electron coupling constants to the $Z$-boson, $M_W$, $M_Z$ and
$\Gamma_Z$ are the $W$ and $Z$-boson masses, and total $Z$ decay width
respectively.

The NU and FC components of the nonstandard cross section,
$\sigma_0^{\rm NSI}=\sigma_0^{\rm NU}+\sigma_0^{\rm FC}$, are given by
\cite{Berezhiani:2001rs}:
\begin{eqnarray}
\sigma_0^{\rm NU}(s) &= & 
\sum_{\alpha=e,\mu,\tau} \frac{G_F^2}{6\pi}s\left[
(\varepsilon_{\alpha \alpha}^{L})^2 +(\varepsilon_{\alpha \alpha}^{R})^2 -
2 (g_{L}\varepsilon_{\alpha \alpha}^{L} + g_{R}\varepsilon_{\alpha \alpha}^{R}) 
\frac{M^2_Z (s- M^2_Z)}{ (s-M^2_Z)^2 +
(M_Z \Gamma_Z)^2}\right] \nonumber \\
&& +\frac{G_F^2}{\pi} \varepsilon_{ee}^{L}M^2_W\left[ 
\frac{(s + M^2_W)^2}{s^2}
\log\left( 
\frac{s + M^2_W}{M^2_W}\right) -
\frac{M^2_W}{s} -\frac32\right] \,, \\
\sigma_0^{\rm FC}(s) &= &  
\sum_{\alpha \ne \beta=e,\mu,\tau} \frac{G_F^2}{6\pi}s\left[
(\varepsilon_{\alpha \beta }^{L})^2 +(\varepsilon_{\alpha \beta}^{R})^2\right] \,.
\label{eq:fccross}
\end{eqnarray}

\subsection{LSND and reactors cross sections}
\label{sec:lsnd-reactors-cross}

The differential cross section for $\nu_e e$ scattering processes in
the presence of NSI can be written as
\begin{eqnarray}
\label{cross-section}
\lefteqn{{{\rm d}\sigma^{\rm theo}_{\rm LSND} \over {\rm d}T}= {2 G_F^2 m_e \over \pi} [ (\tilde 
g_L^2+\sum_{\alpha \neq e}
|\varepsilon_{\alpha e}^{L}|^2)+{} } \nonumber\\ & & {}+
(\tilde g_R^2+\sum_{\alpha \neq e}
|\varepsilon_{\alpha e}^{R}|^2)\left(1-{T \over E_{\nu}}\right)^2-
(\tilde g_L \tilde g_R+ \sum_{\alpha \neq e}|\varepsilon_{\alpha e}^{L}||
\varepsilon_{\alpha e}^{R}|)m_e {T \over E^2_{\nu}}],
\end{eqnarray}
where $T$ is the electron recoil energy, $m_e$ is the electron mass
and $E_\nu$ is the incident neutrino energy. The effective couplings
$\tilde g_{R,L}$ are given as $\tilde g_R=g_R+\varepsilon_{e e}^{R}$
and $\tilde g_L=g_L+\varepsilon_{e e}^{L}$.
For the case of reactors, we have to exchange $L$ by $R$ and
vice-versa.

\subsection{CHARM II cross section}
\label{sec:charm-ii-cross}

In the presence of NSI the differential $\nu_\mu e\to\nu_\alpha e$
cross section relevant for the case of the CHARM II experiment is
given as
\begin{equation}
\frac{{\rm d}\sigma^{\rm theo}_{\rm CHARM}}{{\rm d}y}=\frac{2G_F^2m_e}{\pi}E_\nu\left[\left(\tilde g_{L}^2+\sum_{\alpha\ne\mu}|\varepsilon_{\alpha\mu}^{L}|^2\right)+\left(\tilde g_{R}^2+\sum_{\alpha\ne\mu}|\varepsilon_{\alpha\mu}^{R}|^2\right)(1-y)^2\right]
\label{eq:charm}
\end{equation}
where $\tilde g_{L,R}=g_{L,R}+\varepsilon^{L,R}_{\mu\mu}$ and
$y=(1-\mbox{cos}\theta^*)/2$ is called the inelasticity. Here
$\theta^*$ is the center of mass scattering angle. For the
anti-neutrino case, we simply have to exchange $L$ by $R$ and
vice-versa.

\section{The data}
\label{sec:data}

In what follows we will mainly focus on the effect of the six flavor
conserving non-standard interactions in the above
processes. 
Generalizing to include also the six flavor changing NSI
parameters is straightforward but technically more complex and
somewhat less motivated in view of strong, albeit indirect, bounds
that follow from searches for lepton flavor violation.
Let us now first briefly describe the relevant data used in our global
analysis.

\subsection{The LEP data}
\label{sec:lep-data}

Neutrino-electron NSI will contribute to the cross section of the
interaction $e^+ e^- \to \nu \bar \nu \gamma$ by increasing or
decreasing the expected number of events. The best data on such
interaction has been collected by the four LEP experiments: OPAL,
ALEPH, L3 and
DELPHI~\cite{Barate:1997ue,Barate:1998ci,Heister:2002ut,Abreu:2000vk,Acciarri:1997dq,Acciarri:1998hb,Acciarri:1999kp,Ackerstaff:1997ze,Abbiendi:1998yu,Abbiendi:2000hh}.
The reported measurements are compiled in Table
\ref{Lep2Data}~\cite{Hirsch:2002uv}.
The center of mass energy and luminosity for each of the four LEP
experiments are given in the second and third columns of Table
\ref{Lep2Data}.  The background subtracted experimental cross sections
and the Monte Carlo expectations are given in picobarns in columns
four and five respectively. Column six reports the number of events
observed after background subtraction. The efficiency $\epsilon$ is
given in column seven and finally, the last two columns report the
kinematical cuts: $x = E_\gamma / E_{beam}$, $x_T=x\sin\theta_\gamma$
with $\theta_\gamma$ the angle between the photon momentum and the
beam direction, and $y=\cos\theta_\gamma$.
As in \cite{Hirsch:2002uv}, we have found that our calculation for the
Standard Model LEP cross section, Eq.~(\ref{eq:sm}) without including
the effects of the NSI, disagrees with the Monte Carlo results quoted
by the four collaborations. This might be due to additional specific
experimental cuts beside the ones quoted in the last two columns in
Table~\ref{Lep2Data}.  Such disagreements are included as an
additional theoretical uncertainty which we have added in quadrature
in the calculation of our errors.

\begin{table}[p] 
\caption{Summary of the ALEPH, DELPHI, L3 and OPAL experimental data, 
collected above the $W^+W^-$ production threshold. Wherever a double 
error is listed, the first is statistical and the second is systematic.}
\label{Lep2Data} {\scriptsize
\begin{ruledtabular}
\begin{tabular}{ccccccccc}
  & $\sqrt{s}$ (GeV) & $\cal L\,$(pb$^{-1}$) &
  $\sigma^{\rm mes}$ (pb) & $\sigma^{\rm MC}$ (pb) &N$_{\rm obs}$
& $\epsilon\,(\%)$ & $E_\gamma$ (GeV) & $|y|$  \\ 
\hline\hline
ALEPH
 & 161 & 11.1 &5.3$\pm$0.8$\pm$0.2  & 5.81$\pm$0.03 & 41 & $70$  &$ x_T\geq 0.075$  & $ \leq 0.95$ \\ 

\cite{Barate:1997ue} & 172 & 10.6 &4.7$\pm$0.8$\pm$0.2  & 4.85$\pm$0.04 & 36 & $72$  &$ x_T\geq 0.075$  & $ \leq 0.95$ \\ 
\hline 
\cite{Barate:1998ci}&183&58.5&4.32$\pm$0.31$\pm$0.13&4.15$\pm$0.03&195&$77$&
                                                   $ x_T\geq 0.075$ & $ \leq 0.95$    \\  
\hline 
 & 189 & 173.6 & 3.43$\pm$0.16$\pm$0.06 & 3.48$\pm$0.05 & 484 & &&   \\  
 & 192 &\ 28.9 & 3.47$\pm$0.39$\pm$0.06 & 3.23$\pm$0.05 &\ 81 & &&   \\  
 & 196 &\ 79.9 & 3.03$\pm$0.22$\pm$0.06 & 3.26$\pm$0.05 & 197 & &&   \\  
 \cite{Heister:2002ut}
 & 200 &\ 87.0 & 3.23$\pm$0.21$\pm$0.06 & 3.12$\pm$0.05 & 231 & 81.5& $ x_T\geq 0.075$  & $ \leq 0.95$   \\ 
 & 202 &\ 44.4 & 2.99$\pm$0.29$\pm$0.05 & 3.07$\pm$0.05 & 110 & &&   \\  
 & 205 &\ 79.5 & 2.84$\pm$0.21$\pm$0.05 & 2.93$\pm$0.05 & 182 & &&   \\  
 & 207 & 134.3 & 2.67$\pm$0.16$\pm$0.05
 & 2.80$\pm$0.05 & 292 & &&   \\  
\hline\hline 
DELPHI \cite{Abreu:2000vk}  & && && && &              \\  
HPC
&189& 154.7&1.80$\pm$0.15$\pm$0.14& 1.97 & 146& 51\footnotemark[1] & $x\geq0.06$\ \ & $\leq0.70$\\  

FEMC
&183&\ 49.2&2.33$\pm$0.31$\pm$0.18& 2.08 &\ 65& 54\footnotemark[1] & $x\geq$0.2 & $\geq$0.85 \\ 

FEMC
&189& 157.7&1.89$\pm$0.16$\pm$0.15& 1.94 &\ 155& 50\footnotemark[1] &$x\leq$0.9 &  $\leq$0.98 \\ 

\hline\hline 
L3&161&10.7 &6.75$\pm$0.91$\pm$0.18&6.26$\pm$0.12&57&$80.5$&$\geq10$&$\leq0.73$ \\ 
\cite{Acciarri:1997dq}&&&&&&&$\!\!$ and \qquad\qquad & \\
&172&10.2&6.12$\pm$0.89$\pm$0.14&\ 5.61$\pm$0.10&49&$80.7$&
                                      $E_T\geq6$&  $0.80$--$0.97$\\ 
\hline 
\cite{Acciarri:1998hb} &183 
& 55.3 &5.36$\pm$0.39$\pm$0.10 & 5.62$\pm$0.10 & 195 & $65.4$  & $\geq 5$ & $\leq0.73$   \\  
&&&&&&&$\!\!$ and \qquad\qquad & \\  
 \cite{Acciarri:1999kp} &189 
& 176.4 &5.25$\pm$0.22$\pm$0.07 & 5.29$\pm$0.06 & 572 & $60.8$  & $E_T\geq5$&  $0.81$--$0.97$\\  
\hline\hline 
OPAL
& 130 & 2.3 & $10.0\pm2.3\pm0.4$ & $13.48\pm0.22$\footnotemark[2] & 19 & 81.6 & $x_T>0.05$ & $\leq0.82$ \\
\cite{Ackerstaff:1997ze}&&&&&&&$\!\!$ or \qquad\qquad & \\  
& 136 & 2.59 & $16.3\pm2.8\pm0.7$ & $11.30\pm0.20$\footnotemark[2] & 34 & 79.7 & $x_T>0.1$ & $\leq0.966$ \\
\hline
& 130 & 2.35 & $11.6\pm2.5\pm0.4$ & $14.26\pm0.06\footnotemark[2]$ & 21 & 77.0 & & \\
\cite{Abbiendi:1998yu}
&&&&&&&$\!\!$ $x_T>0.05$ & $\leq0.966$ \\  
& 136 & 3.37 & $14.9\pm2.4\pm0.5$ & $11.95\pm0.07\footnotemark[2]$ & 39 & 77.5 & & \\
\hline
&161 & 9.89 & 5.3$\pm$0.8$\pm$0.2 & 6.49$\pm$0.08\footnotemark[2]&40& $75.2$   
                                                 &  $x_T$$>$$\,0.05$&$\leq0.82$\\ 
\cite{Ackerstaff:1997ze}&&&&&&&$\!\!$ or \qquad\qquad & \\
 &172 &10.28 & 5.5$\pm$0.8$\pm$0.2&  5.53 $\pm$0.08\footnotemark[2]&45& $77.9$ & $x_T$$>$$\,0.1\,$ &  $\leq0.966$\\ 
\hline 
\cite{Abbiendi:1998yu}
 &183 &54.5 & 4.71$\pm$ 0.34$\pm$0.16 & 4.98$\pm$0.02\footnotemark[2] & 191 & $74.2 $  & 
  $x_T$$>$$\,0.05$&  $\leq0.966$ \\  
\hline 
\cite{Abbiendi:2000hh}
&189 &177.3 & 4.35$\pm$0.17$\pm$0.09&4.66$\pm$0.03 & 643 & $82.1 $  &  $x_T$$>$$\,0.05$&  $\leq0.966$ \\
\end{tabular}
\end{ruledtabular} }
\footnotetext[1]{Estimated from the Monte Carlo cross sections and the
expected numbers of events.}
\footnotetext[2]{Calculated from the expected number of events as
  predicted by  the KORALZ event generator.}
\label{tab:lep}
\end{table}

In total the four LEP experiments lead to 25 observables. Because of
the small systematic error they have, we can assume that all of them
are independent with no correlation between them.

\subsection{The LSND and reactors data}
\label{sec:lsnd-reactors-data}

The best measurements of the cross section for the $\nu_ee$ and $\bar
\nu_ee$ scattering processes have been performed in terrestrial
experiments.  The cross section for the elastic scattering interaction
$\nu_e +e^- \to \nu_e + e^-$ was measured by the Liquid Scintillator
Neutrino Detector (LSND) using a $\mu^+$ decay-at-rest $\nu_e$ beam at
the Los Alamos Neutron Science Center. The detector is an
approximately cylindrical tank containing 167 tons of liquid
scintillator and viewed by 1220 photomultiplier tubes. The final
neutrino-electron cross section is reported in Table \ref{nu:scat}~
\cite{Auerbach:2001wg}.
 
\begin{table}
\centering
\caption{Experimental measurements of the $\nu_ee$ and $\bar \nu_ee$
  scattering cross sections}\vskip .2cm
\label{nu:scat}
\begin{tabular}{cccc}
\hline\hline
Experiment & Energy range (MeV) & Events & Measurement \\
\hline 
LSND $\nu_e-e$   & 10-50  & 191 &
$\sigma=[10.1\pm1.5]\times 
E_{\nu_e}({\rm MeV})\times10^{-45} {\rm cm}^2$   \\[.1cm]
Irvine $\bar{\nu}_e-e$  & 1.5- 3.0  & 381 &
$\sigma=[0.86\pm0.25]\times \sigma_{V-A}$   \\[.1cm]
Irvine $\bar{\nu}_e-e$   & 3.0- 4.5 & 77 &
$\sigma=[1.7\pm0.44]\times \sigma_{V-A}$   \\[.1cm]
Rovno $\bar{\nu}_e-e$  & 0.6 - 2.0 & 41 &
$\sigma=(1.26\pm 0.62)\times10^{-44} {\rm cm}^2 / {\rm fission}$ \\[.1cm]
 MUNU $\bar{\nu}_e-e$  & 0.7 - 2.0  & 68 &
$1.07 \pm 0.34$ events day~$^{-1}$  \\[.1cm]
\hline\hline
\end{tabular}
\end{table}

The Irvine \cite{Reines:1976pv}, the most recent
MUNU~\cite{Daraktchieva:2003dr} and the Rovno~\cite{Derbin:1993wy}
experiments have measured the $\bar\nu_ee$ scattering by using
neutrinos from reactors. The measured cross section is also reported
in Table \ref{nu:scat}. We also quoted the number of events for each
experiment in column three and the recoil electron energy range in
column two.

\subsection{The CHARM II data}
\label{sec:charm-ii-data}

The CHARM collaboration used a massive 692 ton target calorimeter
followed by a muon spectrometer to detect the $\nu_\mu + e \to \nu_\mu
+ e $ and $\bar \nu_\mu +e \to \bar \nu_\mu+ e$ scattering processes.
The neutrinos were produced by a 450 GeV proton beam accelerated in
the Super Proton Synchrotron (SPS) for 2.5$\times 10^{19}$ protons on
target. Approximately 10$^8$ neutrino interactions occurred in the
detector. Data collected from 1987-1991 were 2677 $\pm 82$ events for
reaction $\nu_\mu + e \to \nu_\mu + e $ and 2752 $\pm 88$ events in
the $\bar \nu$ beam \cite{Vilain:1994qy}.  There was a neutrino
contamination, of approximately $10\% $ of the flux, in the
muon-antineutrino electron scattering.

The CHARM collaboration used these data to determine the values of the
SM $g_A$ and $g_V$ coupling constants. Because of the quadratic
dependence on the coupling constants in the cross section formula
given in Eq.~(\ref{eq:charm}) there is a well known fourfold ambiguity
in the determination of $g_V$ and $ g_A$. A similar ambiguity in
determining $g^e_V$ and $ g^e_A$ has been removed
in~\cite{Vilain:1994qy} by combining the $\nu_e e $ and $\bar \nu_e e$
scattering data obtained by the CHARM detector with the
forward-backward asymmetry ($A_{FB}$) in the $e^+ e^- \to e^+ e^-$
scattering at LEP~\cite{collaborations1992adl}.
Here we will obtain a similar result in the context of constraining
neutrino NSIs but using the ``neutrino counting'' LEP data.

\section{The $\chi^2$ analysis}
\label{sec:chi2-analysis}

Once we have defined in the previous sections the cross sections for
each of the processes under consideration, and we have introduced all the
experimental measurements relevant for our analysis, we proceed to
perform a $\chi^2$ analysis.

For the LEP data, we can obtain a theoretical estimate of the expected
number of events $N_i^{\rm theo}$ for each of the 25 observables by
using Eq.~(\ref{eq:sm}). The integration of the cross section has been
performed with the experimental cuts reported in last two columns of
Table \ref{Lep2Data}.  We have used the reported luminosity ($\cal L$)
and efficiency ($\epsilon$) for each experiment.

We define the corresponding $\chi^2$ function as
\begin{equation}
\chi^2_{\rm LEP}=\sum_{i=1}^{25}\frac{(N_i^{\rm theo}-N_i^{\rm obs})^2}{\Delta_i^2}\,
\end{equation}
where $N_i^{\rm obs}$ is reported in Table \ref{Lep2Data} and
$\Delta_i$ is the corresponding error.  In the SM limit, our cross
section computations agree within $8\%$ with the LEP Monte Carlo
results, except for L3, where we have found up to a $20\%$
discrepancy.  Therefore, we have allowed for an extra $10\%$
theoretical systematic error added in quadratures to all LEP
experiments~\footnote{We have also found that different error
  assignment prescriptions do not substantially affect the results.}.
We have neglected all correlations since statistical and systematic
errors are small.

For the $\nu_ee$ and $\bar \nu_ee$ scattering processes, we define the
$\chi^2$ as
\begin{equation} 
\chi^2_{\nu_ee} =   
\sum_i\frac{(\sigma_i^{\rm theo}-\sigma_i^{\rm exp})^2}{\Delta_i^2}\,
\end{equation} 
where the $\sigma_i^{\rm exp}$ are given by the experimental
measurements and $\Delta_i$ are the corresponding errors reported in
Table \ref{nu:scat}, while $\sigma_i^{\rm theo}=\sigma_i^{\rm
  SM}+\sigma_i^{\rm NSI}$ are the theoretical expectations considering
the effects of NSI calculated via Eq. (\ref{cross-section}).  Details
of the analysis for LSND and reactor experiments constraining NSI in
$\nu_ee$ and $\bar\nu_ee$ scattering have already been given in
Ref.~\cite{Barranco:2005ps}.

For CHARM II we calculated the number of events using the cross
section from Eq. (\ref{eq:charm}).  With the NSI parameters fixed to
zero we defined a normalization constant to reproduce the number of
events reported by the CHARM II collaboration.

For CHARM II data we have used
\begin{equation}
\chi^2_{\rm  CHARM}=\sum_{i=1}^2\frac{(N_i^{\rm theo}-N_i^{\rm obs})^2}{\Delta_i^2}\,
\end{equation}
where one observable stands for $\nu_\mu e$ scattering and the other
one for $\bar \nu_\mu e$.

The  global $\chi^2$ is simply the sum of the individual ones,
\begin{equation}
\chi^2_{\rm TOT}=\chi^2_{\rm LEP}+\chi^2_{\nu_e e}+\chi^2_{\rm CHARM} \,.
\end{equation}

\section{Results}
\label{sec:results}

The cross section for $e^+e^- \to \bar \nu \nu \gamma$ including NSI
is sensitive to all twelve $\varepsilon_{\alpha \beta}^{L,R}$
parameters. 
On the one hand the scattering interactions $\nu_ee$ and $\bar
\nu_ee$ are sensitive to six parameters: $\varepsilon_{e
  \alpha}^{L,R}$, with $\alpha=e,\mu,\tau$. 
On the other hand the elastic scatterings $\nu_\mu e$ and $\bar
\nu_\mu e$ are sensitive to the other six parameters:
$\varepsilon_{\mu \alpha}^{L,R}$.

In order to obtain constraints on the relevant NSI parameters, we
first follow the most popular approach adopted by the majority of
authors~\cite{Berezhiani:2001rs,Davidson:2003ha}.
It consists on varying only one parameter at-a-time and fixing the
remaining parameters to zero.  This way we obtain bounds on the twelve
NSI parameters.
However such one-parameter-at-a-time analysis is fragile and might
miss potential cancellations in the determination of the restrictions
upon NSI strengths.

As a second step, we assume that the new physics induces mainly
flavor-conserving, effective NU neutral current interactions, so that
the only relevant parameters are the six $\varepsilon _{\alpha
  \alpha}^{L,R}$, $\alpha=e,\mu, \tau$. This is reasonable in view of
the relatively strong bounds on lepton flavor violating processes.

\subsection{One parameter at-a-time}
\label{sec:one-parameter-at}

In this section we constrain the neutrino-electron NSI parameters
varying them one parameter at-a-time. Because each cross section is
sensitive to different parameters, depending on the parameter under
consideration, the number of total observables used in the analysis
will change.  Table \ref{tab:booktabs} shows the $\chi_{min}^2$, the
number of degrees of freedom (d.o.f), the ratio $\chi_{min}^2/$d.o.f,
the allowed range for each of the twelve parameters obtained by our
$\chi^2$ analysis and the limits obtained by previous analyses.

\begin{table}[htbp]
   \centering
   \caption{Constrains of neutrino-electron NSI parameters by varying
     only one parameter at-a-time. Improvements are obtained
     compared with previous analyses. Read text for details.}\vskip .2cm
   \begin{ruledtabular}
   \begin{tabular}{ccccc}
  d.o.f.   & $\chi^2_{min}$  & $\chi^2_{min}/$d.o.f. & 90\% C.L. Allowed Region & Previous Limit~\cite{Davidson:2003ha,Barranco:2005ps} \\
      \hline
       29   & 24.8482 & 0.8568 & $-0.03<\varepsilon_{ee}^L<0.08$ & $-0.05<\varepsilon_{ee}^L<0.1$ ($\nu_ee$)\\
      29   & 22.4742 & 0.7750 & $0.004<\varepsilon_{ee}^R<0.151$ & $-0.04<\varepsilon_{ee}^R<0.14$ ($\nu_e e$)\\
      26   & 22.1308 & 0.8512 & $|\varepsilon_{\mu\mu}^L|<0.03$ & $|\varepsilon_{\mu\mu}^L|<0.03$ ($\nu_\mu e$)\\
      26   & 22.1315 & 0.8512 & $|\varepsilon_{\mu\mu}^R|<0.03$ & $|\varepsilon_{\mu\mu}^R|<0.03$ ($\nu_\mu e$)\\
      24   & 21.8927 & 0.9122 & $-0.46<\varepsilon_{\tau\tau}^L<0.24$ & $-0.6<\varepsilon_{\tau\tau}^L<0.4$ ($e^+e^-\to \nu\nu\gamma$)\\
      24   & 21.9072 & 0.9128 & $-0.25<\varepsilon_{\tau\tau}^R<0.43$ & $-0.4<\varepsilon_{\tau\tau}^R<0.6$ ($e^+e^-\to \nu\nu\gamma$)\\
\hline\hline
      31   & 22.8752 & 0.7379 & $|\varepsilon_{e\mu}^L|<0.13$ & 
$|\varepsilon_{e\mu}^L|\simeq5\times10^{-4}$ ($\mu\to 3e$) \\
      31   & 24.9885 & 0.8061 & $|\varepsilon_{e\mu}^R|<0.13$ & $|\varepsilon_{e\mu}^R|\simeq 5\times10^{-4}$  ($\mu\to 3e$) \\
      29   & 22.3062 & 0.7692 & $|\varepsilon_{e\tau}^L|<0.33$ & $|\varepsilon_{e\tau}^L|<0.4$ ($\nu_ee$)\\
      29   & 22.2107 & 0.7659 & $0.05<|\varepsilon_{e\tau}^R|<0.28$ & $|\varepsilon_{e\tau}^R|<0.27$ ($\nu_ee$)\\
      26   & 22.1308 & 0.8512 & $|\varepsilon_{\mu\tau}^L|<0.1$ & $|\varepsilon_{\mu\tau}^L|<0.1$ ($\nu_\mu e$) \\
      26   & 22.1312 & 0.8512 & $|\varepsilon_{\mu\tau}^R|<0.1$ & $|\varepsilon_{\mu\tau}^R|<0.1$ ($\nu_\mu e$) \\
   \end{tabular}
   \end{ruledtabular}
   \label{tab:booktabs}
\end{table}

One sees how the inclusion of the LEP data leads to an improvement in
the constraints for most of the NU NSI parameters.
For example, from the last column in Table \ref{tab:booktabs} one can see
how previous constraints on $\varepsilon_{e \alpha}^{L,R}$ coming from
LSND and reactor data~\cite{Davidson:2003ha,Barranco:2005ps} are now
superseded. Our analysis also improves previous constraints on
$\varepsilon_{\tau \tau}^{L,R}$.
The inclusion of LEP data also improves the limits for
$\varepsilon_{e\tau}^{L,R}$. Note that a nonzero
$\varepsilon_{e\tau}^R$ is favored in this analysis, though this has no
statistical significance. We can also see that the ratio
$\chi^2_{min}/$d.o.f. is close to unity in the majority of the cases,
meaning that the $\chi^2$ is a good statistical indicator.

Note, however, that there is no improvement in the constraints for the
parameters $\varepsilon_{\mu\alpha}^{L,R}$, since these are dominated
by the CHARM II data and the restrictions from $\mu\to3e$ for the NU
and FC non-standard neutrino interactions,
respectively~\cite{Davidson:2003ha}.

Here a comment on FC NSI is in order. Clearly, if
there are FC NSI on neutrinos one expects, by SU(2) gauge symmetry,
that these will induce also FC on charged leptons, which are rather
strongly constrained by the non-observation of the corresponding LFV
processes such as $\mu\to e\gamma$, $\mu\to 3e$, $\mu\to e$ conversion
in nuclei, $\tau\to\mu\bar{e}e$, $\tau\to\mu\rho$, etc. Indeed, given
the existence of the effective NSI operators one obtains, by
``dressing'' with weak gauge-boson exchange, a corresponding effective
NSI operator involving only charged leptons.  However the loop
diverges logarithmically.  In this case a precise prescription must be
given in order to estimate the corrections since the effective
interactions are nonrenormalizable, and therefore there will be a
dependence on the cuttoff scale $\Lambda$ at which the theory is
supposed to be renormalizable~\cite{Davidson:2003ha}.
While the corresponding logarithmic terms can be rigurously computed
when the physics producing the NSI lies at a large scale, it is
certainly not so when it lies at a relatively low scale. The latter is
precisely the case which is most relevant phenomenologically, for
example, schemes like the extended seesaw, broken R-parity or radiative
models of neutrino
mass~\cite{zee:1980ai,babu:1988ki,AristizabalSierra:2006gb}. In these
cases there is no model-independent way to rigurously compute the
magnitude of the induced NSI among charged leptons in terms of that
among neutrinos. 

It follows that so far NSI involving neutrinos are not strongly
constrained, hence the importance of the constraints reported in Table
\ref{tab:booktabs}: in contrast with LFV constraints these are robust.

Before closing this section let us mention that constraints coming
from solar~\cite{Miranda:2004nb,Guzzo:2004ue}, 
atmospheric~\cite{gonzalez-garcia:1998hj,Fornengo:2001pm,Friedland:2004ah},
and MINOS~\cite{Friedland:2006pi} data can not be directly compared
with the bounds obtained here since those do not probe directly the
NSI parameters but only a combination of them which effectively
affects neutrino propagation in matter. For example, for the solar
case, the relevant quantities, $\varepsilon$ and $\varepsilon'$ are
two effective parameters which, for $\varepsilon_{\alpha\mu}^{P} \sim
0$, are related with the {\it vectorial} couplings by:
\begin{equation}
    \varepsilon = - \sin\theta_{23}\,\varepsilon_{e\tau}^{V} \qquad
    \varepsilon^\prime = \sin^2\theta_{23}\,\varepsilon_{\tau\tau}^{V} -
    \varepsilon_{ee}^{V} \,,
    \label{eff-coup}
\end{equation}
with $\varepsilon_{\alpha\beta}^{V}=
\varepsilon_{\alpha\beta}^{L}+\varepsilon_{\alpha\beta}^{R}$. Moreover,
instead of just the NSI with electrons, one should in general take
into account also the possible non-standard interactions with $u$ and
$d$~-type quarks so that the effective NSI parameter becomes:
$\varepsilon_{\alpha\beta}^{P}\equiv
\sum_{f=u,d,e}\varepsilon_{\alpha\beta}^{f P}n_f/n_e$ (with $n_f$ the
density of fermions in the medium). This leaves substantial freedom to
new NSI-induced effects.

\subsection{Flavor-conserving non-universal NSI}
\label{sec:non-universal-nsi}

Barring a theory of flavor, there is no guidance on the structure of
the effective four-Fermi weak interaction.  Generically new physics
will lead to the violation of universality as well as the violation of
leptonic flavor.
In view of the relatively strong constraints on lepton flavor
violating processes it is reasonable, as already mentioned, to first
consider the case of purely flavor-conserving non-standard
interactions, in general non-universal.
In this case the only relevant parameters for our analysis are the
six NU $\varepsilon_{\alpha \alpha}^{L,R}$.  

\begin{figure}
\includegraphics[height=6cm,width=0.32\textwidth]{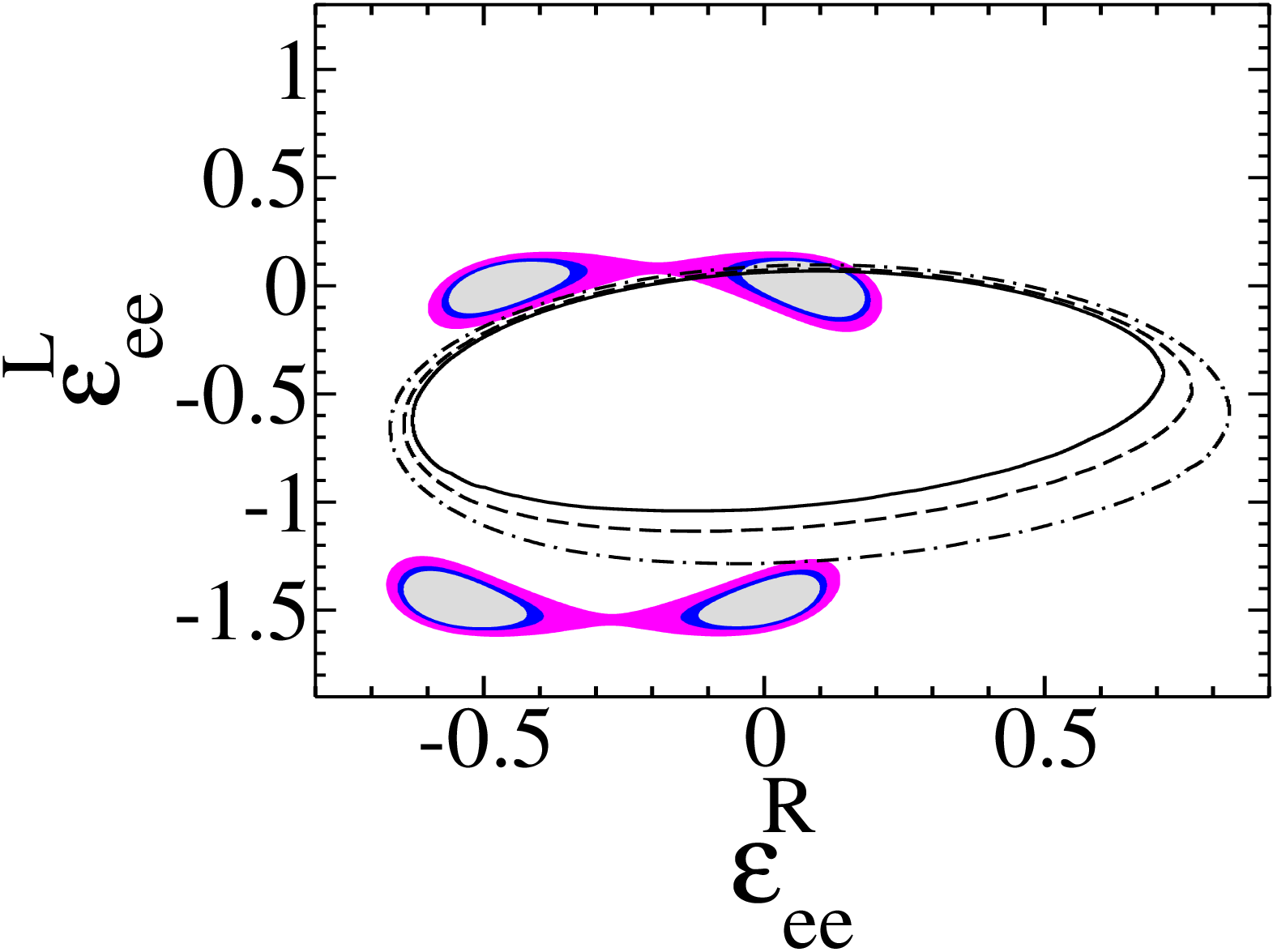}
\includegraphics[height=6cm,width=0.32\textwidth]{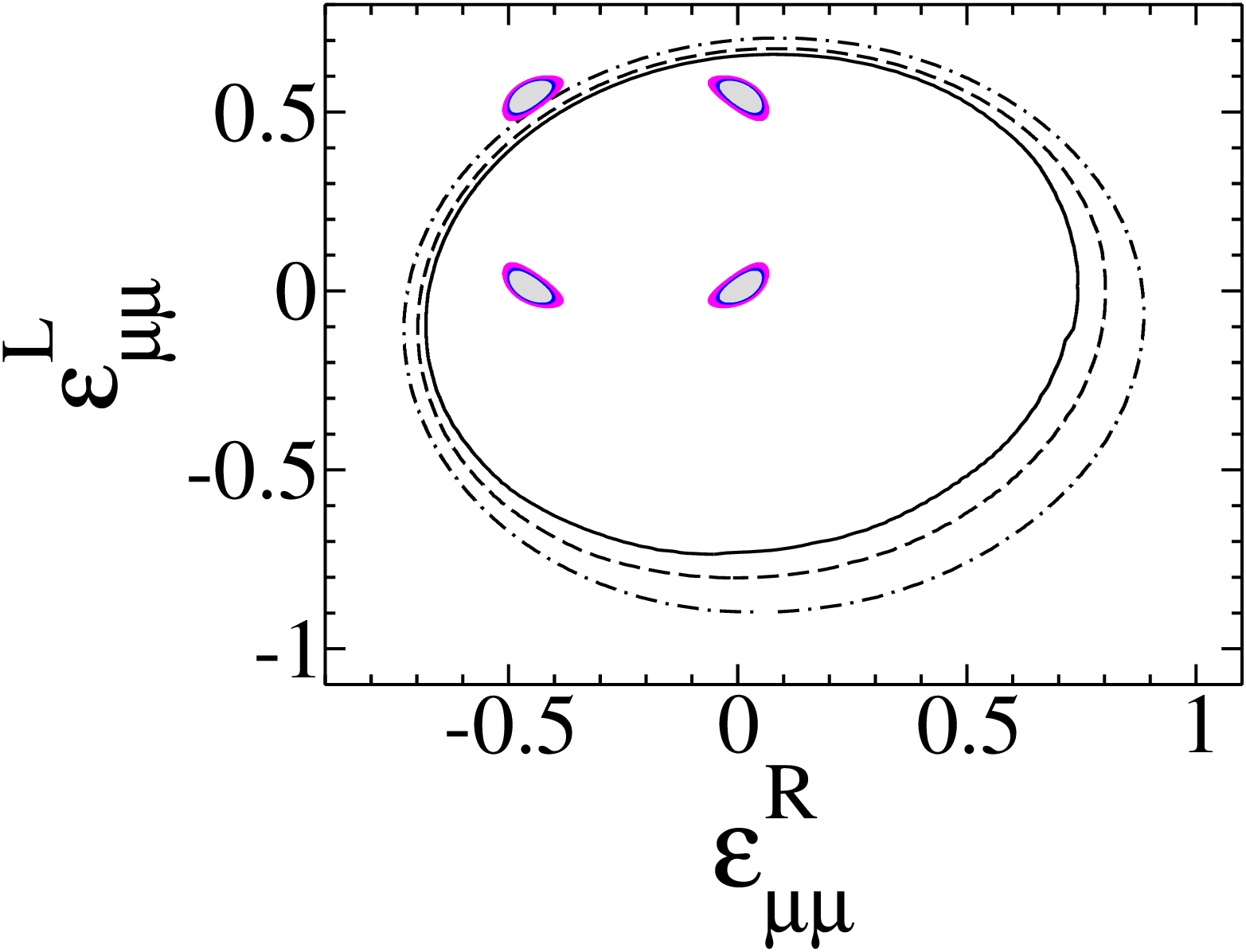}
\includegraphics[height=6cm,width=0.32\textwidth]{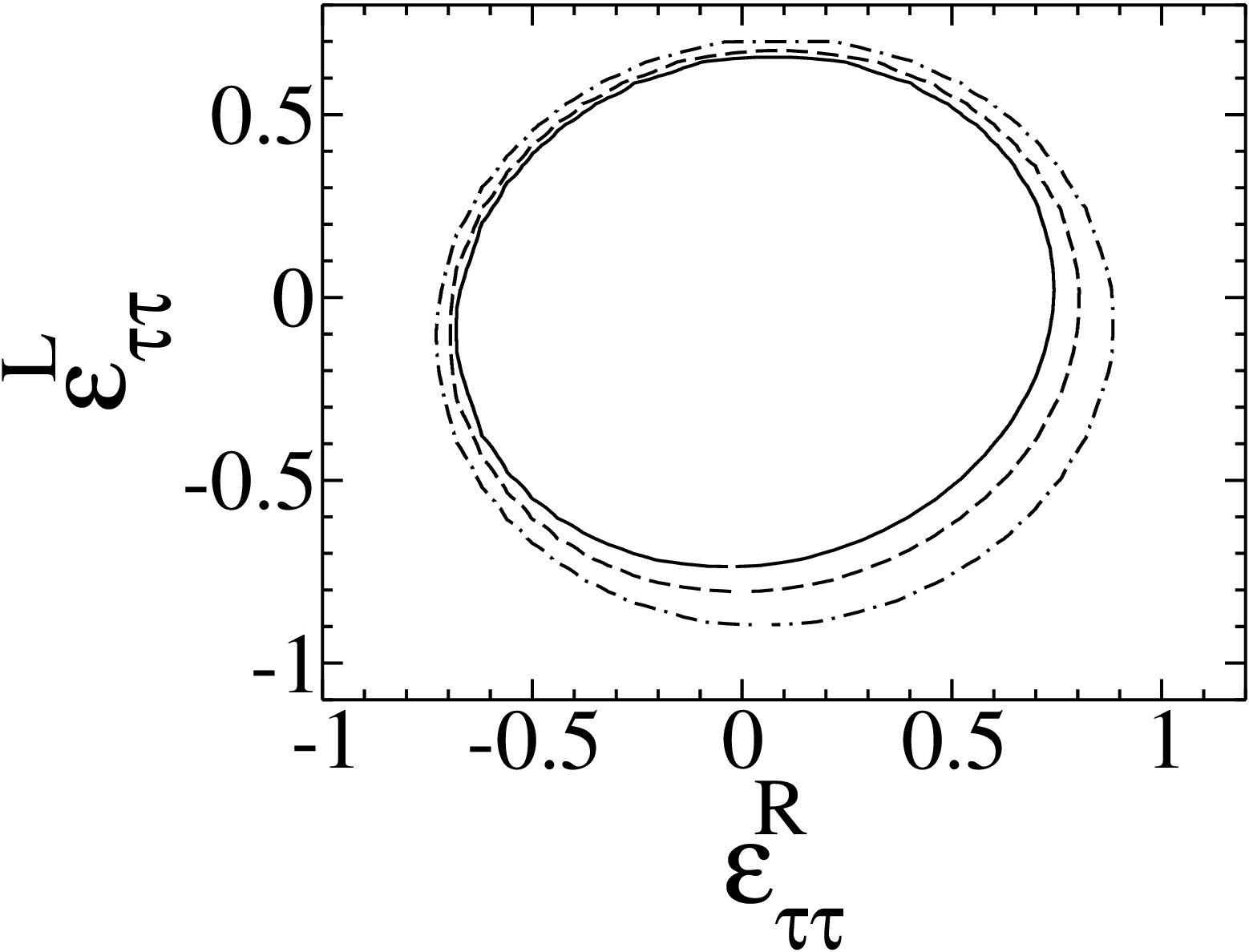}
\caption{
  Flavor conserving NSI allowed by each experiment in our analysis.
  The colored (shaded) regions in the first and second panel show the
  allowed 90 \%, 95 \% and 99 \% C.L. regions from the $\nu_e e$ and
  CHARM II data, respectively. For the LEP data, we take the
  projection of the six free parameters, over the two displayed NSI
  parameters in each case.  In this case, solid, dashed and dot-dashed
  lines show the 90 \%, 95 \% and 99 \% CL allowed regions
  respectively.
}
\label{fig:nonuniversal-1}
\end{figure}

LSND and neutrino reactor data have been used previously in order to
constrain the NU NSI parameters~\cite{Barranco:2005ps}.
It was noted that due to the nature of the elastic neutrino-electron
scattering there is a fourfold ambiguity in the determination of the
NSI parameters. The same happens when the analysis is performed for
the non-universal parameters entering the $\nu_\mu e$ and $\bar\nu_\mu
e$ scattering in the CHARM experiment.
This fourfold ambiguity is clearly seen in the first two panels in
Fig.~\ref{fig:nonuniversal-1}.  The colored regions in the first panel
show the two-dimensional projections in the
$\varepsilon_{ee}^{L}-\varepsilon_{ee}^{R}$ plane arising from the
LSND and reactor data, while the corresponding restriction on the
relevant parameters $\varepsilon_{\mu \mu}^{L,R}$ arising from the
CHARM experiment is displayed in the second panel of
Fig.~\ref{fig:nonuniversal-1}.
Finally, the third panel shows the projection of the constraints
coming from the LEP data only on the parameters $\varepsilon_{\tau
  \tau}^{L,R}$.
The dashed ellipses in the panels indicate the projections of the
constraints following from LEP data only.

These plots clearly indicate the complementarity between the
``inclusive'' LEP $e^+ e^- \to \nu \bar\nu \gamma$ data with those
from reactors \& LSND in constraining electron-type NSIs. Similar
complementarity holds between LEP data and CHARM II data when
constraining muon-type NSIs.

\begin{figure}
\includegraphics[width=0.63\textwidth]{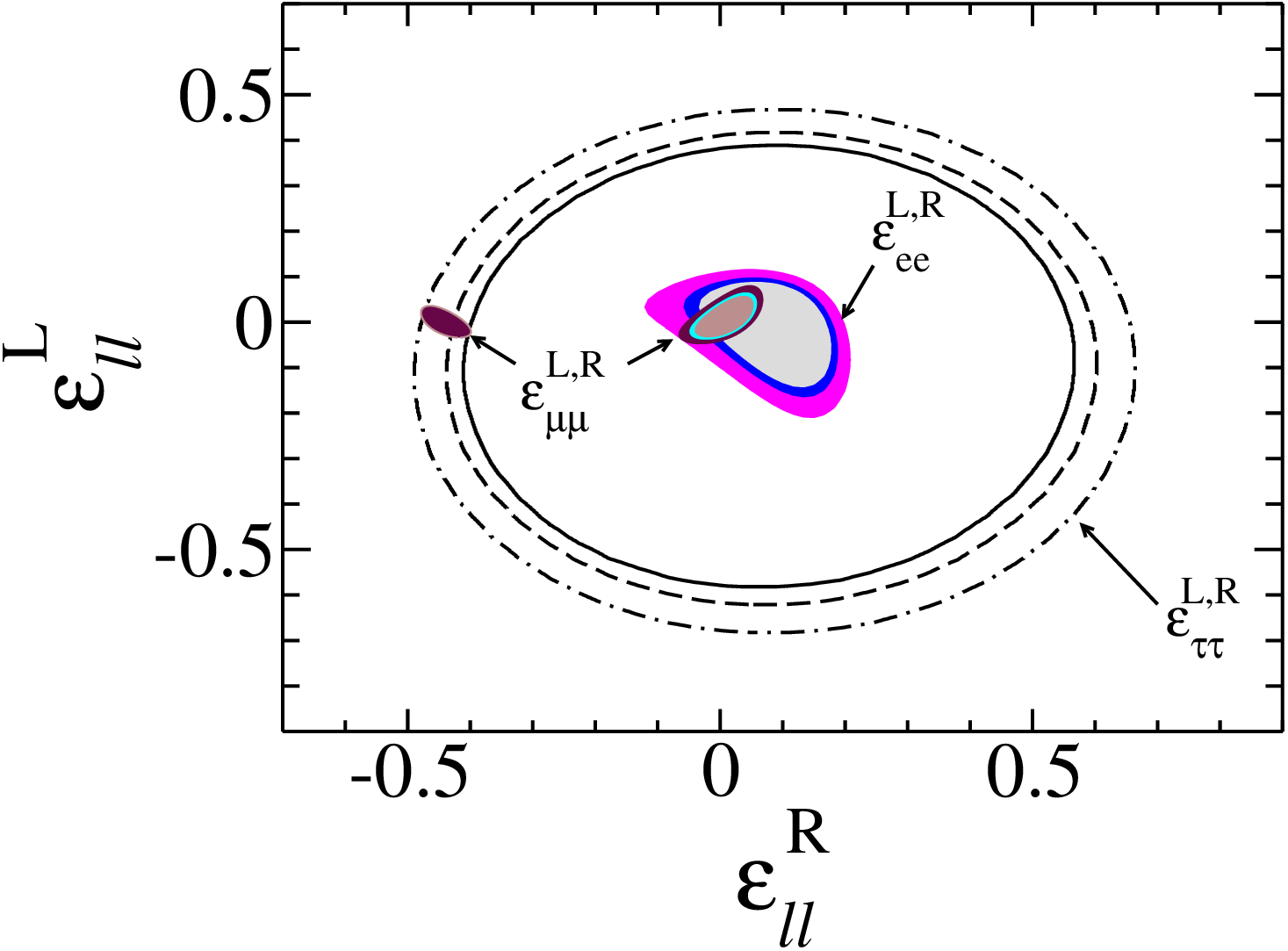}
\caption{
Global analysis results for the flavor conserving NSI
  (details in Table~\ref{tab:booktabs2}).  The plot shows the regions
  allowed at 90 \%, 95 \% and 99 \% C.L. arising from the marginalization of
  the six parameters over different flavors $\varepsilon^{L,R}_{ll}$,
  for $l=e,\mu$ and $\tau$, as indicated. For each flavor the allowed
  region is unique, except for the case of
  $\varepsilon^{L,R}_{\mu\mu}$, where there is a second solution,
  which is allowed only at the 99\%\CL}
\label{fig:nonuniversal-2}
\end{figure}

In the global analysis where $\chi_{\rm TOT}^2$ is the addition of the
LEP, CHARM, LSND and reactor pieces, one clearly sees how the above
fourfold ambiguities are eliminated. This is illustrated in
Fig.~\ref{fig:nonuniversal-2} where we show the allowed regions that
arise from the global $\chi_{\rm TOT}^2$ after taking a projection
over two parameters. The shaded (colored) areas shows the 90 \%, 95
\%, and 99 \% C.L. allowed regions (corresponding to $\Delta
\chi_{\rm TOT}^2=\Delta\chi^2_{min}+ 4.61,$~$5.99,$~$9.21$
respectively). The constraints derived from this analysis are
collected in Table \ref{tab:booktabs2} and compared with the results
discussed in the previous section.  One can see that the interplay
between the different experiments, namely, the combination of the LEP
neutrino counting results with the remaining data, plays a crucial
role in providing constraints almost as stringent as in those obtained
in a one-at-a-time analysis. Needless to say the global analysis
establishes the robustness of these constraints since we are allowing
all the six parameters to vary.

\begin{table}[htbp]
   \centering
   \caption{Constraints for the flavor conserving parameters. We have
   included the data from reactor, CHARM II and LEP experiments and
   allowed the six flavor conserving NSI to be present.
   The global minimum for the
   $\chi^2$ analysis is $\chi^2_{min}=23.13$,
   $\chi^2_{min}/$~d.o.f.~$=1.22$.  
   We show the 90 \% C.L. values
   obtained after taking a projection over two parameters.
   For comparison, we show the constraints for the case in which only
   one parameter is allowed to vary,  and 
   finally, we also compare with previous reported results for the case of 
   one parameter at a time.}
\vskip .4cm
   \begin{ruledtabular}
   \begin{tabular}{lccc}
  & 90\% C.L. Allowed Region & One parameter & Previous limits\\
      \hline
      $\varepsilon_{ee}^L$ &  $-0.14<\varepsilon_{ee}^L<0.09$ & 
         $-0.03<\varepsilon_{ee}^L<0.08$  & $-0.05<\varepsilon_{ee}^L<0.1$  \\
      $\varepsilon_{ee}^R$ &  $-0.03<\varepsilon_{ee}^R<0.18$ & 
   $0.004<\varepsilon_{ee}^R<0.15$  & $0.04<\varepsilon_{ee}^R<0.14$  \\
\hline
      $\varepsilon_{\mu\mu}^L$ & $-0.033<\varepsilon_{\mu\mu}^L<0.055$ 
      & $|\varepsilon_{\mu\mu}^L|<0.03$ & $|\varepsilon_{\mu\mu}^L|<0.03$ \\
      $\varepsilon_{\mu\mu}^R$ & $-0.040<\varepsilon_{\mu\mu}^R<0.053$ 
  & $|\varepsilon_{\mu\mu}^R|<0.03$ & $|\varepsilon_{\mu\mu}^R|<0.03$ \\
\hline
      $\varepsilon_{\tau\tau}^L$ & $-0.6<\varepsilon_{\tau\tau}^L<0.4$  
   &$-0.5<\varepsilon_{\tau\tau}^L<0.2$&$|\varepsilon_{\tau\tau}^L|<0.5$\\
      $\varepsilon_{\tau\tau}^R$ & $-0.4<\varepsilon_{\tau\tau}^R<0.6$ 
   & $-0.3<\varepsilon_{\tau\tau}^R<0.4$&$|\varepsilon_{\tau\tau}^R|<0.5$\\
   \end{tabular}
   \end{ruledtabular}
   \label{tab:booktabs2}
\end{table}

\section{Discussion and Conclusions}
\label{sec:discussion}

We have given a detailed analysis on non-standard neutrino
interactions with electrons following from combining muon and electron
(anti)-neutrino data collected in existing accelerators and reactors,
with the high energy ``neutrino counting'' data from LEP.
Except for $\varepsilon_{\mu\mu}^{L,R}$ and most FC NSIs, the
inclusion of the LEP data within a simple one-parameter-at-a-time
analysis improves upon previous constraints on the flavor-conserving
NSI parameters.

Barring a fundamental theory of flavor, there is no theoretical
guidance on the flavor structure of the NSI that presumably result
from the basic underlying theory producing neutrino masses.
As a result the expected modifications in muon and electron
(anti)-neutrino interactions involve the various components of the
NSIs.
Given this, it is necessary to perform a more general and robust
analysis in which ideally all NSI parameters are allowed to vary
freely. As a first step we have considered the case of non-universal
NSIs. Our results indicate a strong complementarity between the
``neutrino counting'' data and the rest in removing the ambiguous
determination of NSI parameter bounds.
We have obtained unique allowed regions at 90\% and 95\%\CL in NSI
parameter space.
Our improved constraints still leave substantial room for improvement,
posing a big challenge for the next generation of neutrino
experiments.

We thank Martin Hirsch and Arcadi Santamaria for useful comments. This
work has been supported by CONACyT, DGAPA-UNAM, by Spanish Grant No.
FPA2005-01269, and by the Generalitat Valenciana Grant No. ACOMP07/270


\end{document}